\documentclass[prd,twocolumn,nofootinbib]{revtex4-1}
\usepackage{epsfig,psfrag,graphics,color,verbatim}
\usepackage{amsmath}
\def\s#1{\widetilde{#1}}
\def\neut{{\tilde\chi_{1}^0}}
\def\relic{\Omega_{\neut}}

\newcommand{\like}{{\mathcal L}}

\begin{document}

\title{Complementarity of Indirect and Accelerator Dark Matter Searches} 
\author{G.~Bertone\,$^{1}$ D.G.~Cerde\~no\,$^{2}$ M.~Fornasa$^{3,4} \,$ L.~Pieri\,$^{5,6}$ R.~Ruiz de Austri\,$^{7}$ and  R.~Trotta\,$^{8}$}
\affiliation{${^1}$ GRAPPA Institute, University of Amsterdam, Science Park 904, 1090 GL Amsterdam, Netherlands}
\affiliation{${^2}$ Departamento de F\'{\i}sica Te\'{o}rica, and Instituto de F\'{\i}sica Te\'{o}rica UAM/CSIC, Universidad Aut\'{o}noma de Madrid, Cantoblanco, E-28049 Madrid, Spain}
\affiliation{${^3}$ Instituto de Astrof\'{\i}sica de Andaluc\'{\i}a (CSIC), E-18008, Granada, Spain} 
\affiliation{${^4}$ MultiDark fellow} 
\affiliation{${^5}$ Dipartimento di Fisica, Universit\'a di Trento, Via Sommarive 14, I-38123, Trento, Italy}
\affiliation{${^6}$ Istituto Nazionale di Fisica Nucleare, sede di Padova, Via Marzolo 8, 35131 Padova, Italy}
\affiliation{${^7}$ Instituto de F\'isica Corpuscular, IFIC-UV/CSIC, Valencia, Spain} 
\affiliation{${^8}$ Astrophysics Group, Imperial College London, Blackett Laboratory, Prince Consort Road, London SW7 2AZ, United Kingdom} 

\begin{abstract}
In a recent paper, we have shown that even if Supersymmetric particles are found at the Large Hadron Collider (LHC), it will be difficult to prove that they constitute the bulk of the DM in the Universe, while a more convincing case can be made with a combined analysis of LHC and direct detection (DD) data.
Here, we discuss the complementarity of LHC and DM {\it indirect} detection data, focussing in particular on the impact of upper limits on the self-annihilation cross section of Dark Matter particles arising from null searches for gamma-rays from dwarf galaxies with the Fermi LAT satellite. For the specific example in the coannihilation region of a 24-parameters supersymmetric model discussed in the previous paper, these constraints allow in particular to rule out a spurious Wino-like solution, which corresponds to a high self-annihilation cross section and a low relic density, even when astrophysical uncertainties are fully taken into account. We also discuss the impact of upcoming observational data, be them upper limits or actual detection, from the CTA and Planck experiments, and we argue that if New Physics is discovered at the LHC, irrespective of the specific underlying theoretical setup, indirect searches will be useful to break degeneracies in the theoretical parameter space, without making any assumption on the expansion rate of the Universe at the epoch of the Dark Matter freeze-out.   
\end{abstract}

\begin{abstract}
Even if Supersymmetric particles are found at the Large Hadron Collider (LHC), it will be difficult to prove that they constitute the bulk of the Dark Matter (DM) in the Universe using LHC data alone. We study the complementarity of {LHC} and DM {\it indirect} searches, working out explicitly the reconstruction of the DM properties for a specific benchmark model in the coannihilation region of a 24-parameters supersymmetric model. Combining mock high-luminosity {LHC} data with present-day null searches for gamma-rays from dwarf galaxies with the Fermi LAT, we show that current Fermi LAT limits already have the capability of ruling out a spurious Wino-like solution that would survive using LHC data only, thus leading to the correct identification of the cosmological solution. We also demonstrate that upcoming Planck constraints on the reionization history will have a similar constraining power, and discuss the impact of a possible detection of gamma-rays from DM annihilation in Draco with a CTA-like experiment. Our results indicate that indirect searches can be strongly complementary to the LHC in identifying the DM particles, even when astrophysical uncertainties are taken into account.

\end{abstract}

\maketitle

\section{Introduction}
\label{sec:introduction}

One of the most important questions in Dark Matter research (DM) is to identify a possible DM candidate seen at the LHC with the particle responsible for the cosmological relic density. Some of us have recently shown  \cite{Bertone:2010rv} (hereafter Paper I) that a convincing identification of DM particles \cite{Jungman:1995df,Munoz:2003gx,Bertone:2004pz,book,Bertone:2010at} can be achieved with a combined analysis of direct detection and accelerator data. The starting point of Paper I was the simulated response of the Large Hadron Collider (LHC) to a specific benchmark in the coannihilation region of a 24-parameters Minimal Supersymmetric Standard Model (MSSM), and the attempt to reconstruct the properties of the DM particle, in this case the Supersymmetric neutralino, using only these simulated data. As already discussed in Ref. \cite{Baltz:2006fm}, even 300 fb$^{-1}$ of LHC data with a center of mass energy of 14 TeV (i.e. with a dataset that should only become available by 2018 or so, based on current plans), would not allow to identify the neutralino as the sole constituent of the DM in the Universe. Assuming a standard expansion rate at freeze-out, in fact, the posterior probability obtained after imposing the simulated LHC data exhibits multiple modes, corresponding to different neutralino compositions, and a broad peak around the \emph{true} solution. 

We argued in Paper I that a robust and powerful way of breaking the degeneracy in the parameter space, and therefore of identifying the DM particle, is to combine accelerator data with ton-scale direct detection experiments measuring the recoil energy of nuclei struck by DM particles (see Refs. \cite{Munoz:2003gx,Bertone:2004pz,book} and references therein), that should become available over a similar timescale. We demonstrated that a simple {\em Ansatz} on the local density of DM particles (which assumes that the relative abundance of the particles discovered in accelerators with respect to DM is the same locally as their relic density in the Universe) is sufficient to eliminate the spurious mode in the posterior distribution and to constrain the DM properties around the true benchmark value. 
Similarly to the case of accelerators, a single (realistic) direct detection experiment can hardly provide an accurate determination of the DM properties \cite{Green:2007rb,Green:2008rd}, but a second direct detection with a different target would actually allow a much more precise determination of the Weakly Interacting Massive Particle (WIMP) mass \cite{Drees:2008bv}, and if the new target is sensitive to the spin-dependent contribution of the WIMP-nucleus cross section it could even be used to discriminate among WIMP candidates \cite{Bertone:2007xj}. 

Here, we investigate the impact of adding information from {\it indirect}, instead of direct, DM searches. This detection strategy is based on the search for the annihilation or decay products of DM particles, such as high energy photons, neutrinos, and anti-matter \cite{Bertone:2004pz,book}. There are advantages and disadvantages with respect to direct searches. The most obvious advantage is that indirect searches do not require dedicated experiments. Although DM physics has historically played a role in establishing the physics case of experiments such as the Fermi LAT \cite{Fermi,Atwood:2009ez}, HESS \cite{HESS,Aharonian:2005zz} and IceCube \cite{IceCube,Halzen:2010yj}, there is a broad range of other astrophysical motivations that made the construction of these experiments possible. Among the disadvantages, the biggest one arises from the large uncertainties in the predicted annihilation rates (which in turn are a consequence of our poor knowledge of the distribution of DM in the Galaxy and in other astrophysical structures), as well as in the astrophysical backgrounds (see also the discussion in Ref. \cite{Bertone:2010at}). 

Gamma-rays are often considered as ideal messengers for indirect detection studies, since they are not significantly affected by diffusion or energy losses in the local universe. The most promising targets are the Galactic center (GC) \cite{Bergstrom:1997fj,Gondolo:1999ef,Bertone:2002je,Regis:2008ij,Jeltema:2008hf,Hooper:2010mq} and substructures in the Milky Way halo, including dwarf galaxies \cite{Stoehr:2003hf,Strigari:2006rd,Pieri:2007ir,Pieri:2008MNRAS,Pieri:2009AA,Pieri:2009MNRAS,Baltz:2008wd,Martinez:2009jh} and intermediate mass black holes (IMBHs) \cite{Zhao:2005zr,Bertone:2005xz,Bertone:2006nq,Fornasa:2007nr,Aharonian:2008wt,Taoso:2008qz,Bertone:2009kj,Sandick:2011rp}. 

Although it has long been considered as the optimal target, the Galactic center is actually a very problematic region. The first big obstacle to a reliable identification of DM, or at least for the derivation of robust upper limits on the annihilation flux, is represented by the large uncertainties on the DM distribution in a region which is largely dominated by baryons \cite{Merritt:2006mt,Bertone:2005xv,Bertone:2005hw, Pieri:2009je,Pieri:2010PRD}. Furthermore, there are strong and poorly understood astrophysical backgrounds at the GC, that significantly complicate the extraction of a DM signal (see also the discussion in Ref. \cite{Hooper:2011ti}). As for IMBHs, the formation scenarios are not very predictive and rather uncertain, therefore although they might be discovered as a class of objects with identical gamma-ray spectrum and no astrophysical counterparts, the non-detection does not set stringent constraints on the DM particles.

Here, we focus instead on dwarf spheroidal galaxies (dSphs), and analyse the dramatic implications for the reconstruction procedure of DM properties from future LHC data combined with the current Fermi LAT combined analysis of 10 dwarfs \cite{Collaboration:2011wa}. We demonstrate that the spurious solution at low relic density is ruled out by Fermi LAT, if one assumes that the particle found at the LHC makes up all of the DM in the Universe. We also discuss the constraints that should become available in the next few years with the upcoming ground-based gamma-ray experiment CTA, and with a suitable analysis \cite{Galli:2009zc,Slatyer:2009yq,Hutsi:2011vx,Finkbeiner:2011dx,Galli:2011rz} of Planck satellite data. 

The paper is organized as follows: in Section 2 we describe our theoretical setup and the implementation of experimental constraints. In Section 3 we show how the constraining powet of future LHC data can be improved by including current Fermi LAT constraints on the gamma-ray flux from dwarf galaxies, future Planck data probing the reionization history of the Universe and a possible detection of gamma-ray from Draco with CTA.  We conclude in Section 4.  

\section{Theoretical Setup}

\subsection{Benchmark Model}

For concreteness, we will start by defining our theoretical framework. 
In this work we will consider that the MSSM is the correct description for physics beyond the Standard Model and that it also provides a solution to the DM problem in terms of the lightest neutralino, $\s\chi^0_1$.

Neutralinos are physical superpositions of the superpartners of the $B$ and $W$ gauge bosons (Bino and Wino, respectively) and Higgs bosons (Higgsinos) and their phenomenological properties are extremely sensitive to their specific composition. 
In particular, their annihilation cross section in the early Universe (which determines their relic abundance) as well as in DM haloes (which affects the gamma-ray flux that can be observed in indirect DM detection) has an uncertainty of several orders of magnitude depending on the neutralino composition.
It is worth remembering in this sense that neutralinos with a large Wino or Higgsino component are known to have a larger annihilation cross section (and hence a smaller relic density) than those in which the Bino component dominates. 
For this reason, obtaining the neutralino composition in the LHC is a key ingredient for being able to determine whether or not this particle is the main component of the DM.

We follow here the theoretical setup and the notation in Paper I: we adopt a 
minimal supersymmetric extension of the Standard Model (MSSM) with 24 free 
parameters, corresponding to its CP-conserving version. 
The input parameters are the coefficients of the trilinear terms for the 
three generations, the mass terms for gauginos (for which no universality 
assumption is made), right-handed and left-handed squarks and leptons, the 
mass of the pseudoscalar Higgs, the Higgsino mass parameter $\mu$, and 
finally the ratio between the vacuum expectation values of the two Higgs 
bosons $\tan\beta$. We assume a benchmark point corresponding to the 
low-energy extrapolation of model LCC3 as defined in Ref.\,\cite{Baltz:2006fm}. 
This benchmark is representative of SUSY models in the coannihilation region, 
where the lightest neutralino is almost degenerate in mass with the lightest 
stau. 
In this region, coannihilation effects reduce the neutralino relic abundance 
down to values compatible with the results from the WMAP satellite 
\cite{Jarosik:2010iu}, and therefore, the mass difference between the 
neutralino and the lightest stau is a fundamental parameter for the 
reconstruction of the relic density.

For this MSSM model, the structure of the neutralino sector is very characteristic of the MSSM, with the lightest neutralino is Bino-like, the second lightest neutralino is Wino-like and the two heavier neutralinos correspond to Higgsino-like states with a relatively large mass (approximately 460 GeV).
This structure is a consequence of the low-energy hierarchy $M_1<M_2<\mu$ among the gaugino and Higgsino mass parameters at low-energy. 
Because of this, the lightest chargino is also Wino-like, the heaviest corresponding to the charged Higgsino. 

\subsection{Statistical analysis}

In order to constrain the parameters $\mathbf{x}$ described above of our 24-dimensional SUSY 
model, we make use of Bayes' 
theorem
\begin{equation}
p(\mathbf{x}|\mathbf{d})=
\frac{p(\mathbf{d}|\mathbf{x})p(\mathbf{x})}{p(\mathbf{d})},
\label{eqn:Bayesian_theorem}
\end{equation}
which updates the so-called prior probability density $ p(\mathbf{x}) $, 
representing the (lack of) knowledge on the 24-dimensional space before taking into 
account the experimental constraints, $ \mathbf{d} $, to the posterior 
probability function (pdf) $ p(\mathbf{x}|\mathbf{d}) $. The latter is the probability density after the data have been taken into account via the likelihood 
function, $ p(\mathbf{d}|\mathbf{x}) = \like(\mathbf{x})$. Furthermore, on the RHS of 
Eq.\,\eqref{eqn:Bayesian_theorem}, $ p(\mathbf{d}) $ is the Bayesian evidence 
which, in our case, can be dropped since it simply plays the role of a 
normalization constant for the posterior in this context (see 
Ref.~\cite{Trotta:2008qt} for further details). 

The posterior encodes both the information contained in the priors and in 
the experimental constraints, but, ideally, it should be largely independent 
of the choice of priors, so that the posterior inference is dominated by the 
data contained in the likelihood. If some residual dependence on the prior
$ p(\mathcal{\mathbf{x}}) $ remains this should be considered as a sign that 
the experimental data employed are not constraining enough to override 
completely different plausible prior choices and therefore the resulting 
posterior should be interpreted with some care, as it might depend on the 
prior assumptions~\cite{Trotta:2008bp,Roszkowski:2009ye}. For the practical implementation of the Bayesian analysis 
sketched above we employed the \texttt{SuperBayeS} code \cite{deAustri:2006pe,SuperBayeS},  
extending the publicly available version 1.5 to the 
24 dimensions of our SUSY parameter space. To scan in an efficient way the 
SUSY parameter space we have upgraded the MultiNest~\cite{Feroz:2008xx} 
algorithm included in SuperBayeS to the latest MultiNest release (version 2.7), with uniform priors on the pMSSM parameters as in Paper I, to which the reader is referred for full details. 

The set of LHC simulated measurements that we use as constraints in our analysis 
corresponds to that in Table 6 of Ref.\,\cite{Baltz:2006fm}, which assumes an 
integrated luminosity of 300 fb$^{-1}$. Furthermore, as pointed out in 
Ref.\,\cite{Arnowitt:2008bz}, the neutralino-stau mass difference can 
be measured with an accuracy of 20\% with a luminosity of 10 fb$^{-1}$ in
models where the squark masses are much larger than those of the lightest 
chargino and second-lightest neutralino, as is our case. 
We therefore also include a measurement of the neutralino-stau mass 
difference in our likelihood. We summarize in Table \ref{tab:constraints} 
the set of LHC measurements adopted in our likelihood. Each of the 
constraints listed in Table \ref{tab:constraints} is implemented in the 
likelihood as an independent Gaussian distributed measurement around the 
true value $\mu_i$ for that observable, with standard deviation $\sigma_i$, 
as given in Table~\ref{tab:constraints}, i.e. the likelihood from the LHC has the form:
\begin{equation}
\like_\text{LHC}(\mathbf{x}) \propto \prod_i \exp\left(-\frac{1}{2} \frac{(\mu_i - \mu_i(\mathbf{x}))^2}{\sigma_i^2} \right),
\end{equation}
where $ \mu_i(\mathbf{x})$ is the predicted value for the observable at point $\mathbf{x}$ in parameter space.

\begin{table}
\centering
\begin{tabular}{lcc}
\hline
Mass & Benchmark value, $\mu$ \quad  & \quad LHC error, $\sigma$\quad  \\
\hline
$ m(\s\chi^0_1) $ & 139.3 & 14.0 \\
$ m(\s\chi^0_2) $ & 269.4 & 41.0  \\
$ m(\s e_1) $ & 257.3 & 50.0 \\
$ m(\s \mu_1) $ & 257.2 & 50.0 \\
$ m(h) $ & 118.50 & 0.25 \\
$ m(A) $ & 432.4 & 1.5 \\
$ m(\s \tau_1)- m(\s\chi^0_1) $ & 16.4 & 2.0 \\
$ m(\s u_R) $ & 859.4 & 78.0 \\
$ m(\s d_R) $ & 882.5 & 78.0 \\
$ m(\s s_R) $ & 882.5 & 78.0 \\
$ m(\s c_R) $ & 859.4 & 78.0 \\
$ m(\s u_L) $ & 876.6 & 121.0 \\
$ m(\s d_L) $ & 884.6 & 121.0 \\
$ m(\s s_L) $ & 884.6 & 121.0 \\
$ m(\s c_L) $ & 876.6 & 121.0 \\
$ m(\s b_1) $ & 745.1 & 35.0 \\
$ m(\s b_2) $ & 800.7 & 74.0 \\
$ m(\s t_1) $ & 624.9 & 315.0  \\
$ m(\s g) $ & 894.6 & 171.0 \\
$ m(\s e_2) $ & 328.9 & 50.0 \\
$ m(\s \mu_2) $ & 328.8 & 50.0 \\
\hline
\end{tabular}
\caption{Sparticle spectrum (in GeV) for our benchmark SUSY point and relative estimated measurements errors at the LHC (standard deviation $\sigma$).}  
\label{tab:constraints}
\end{table}

\begin{figure}
\epsfig{file=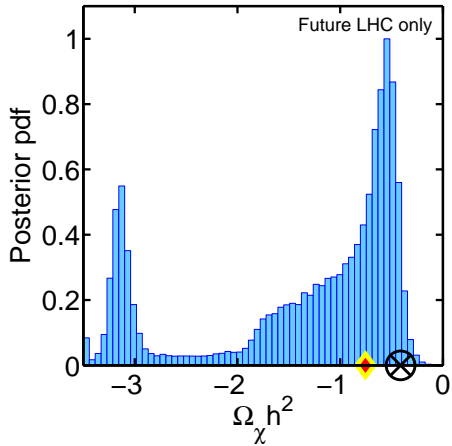,width=0.4\textwidth}
\caption{1-D marginalized posterior probability density function (pdf) for the neutralino relic abundance after LHC hypothetical measurements (given in Table 1) are taken as experimental constraints. The true value is given by the yellow/red diamond.}
\label{pdfLHC}
\end{figure}

After imposing these constraints, the 1D marginalized posterior pdf for the neutralino relic abundance is shown in Fig\,.\ref{pdfLHC}. Despite the good accuracy in the determination of some of the supersymmetric masses (for example, the neutralino mass has a relatively small 10\% uncertainty) and  especially the mass difference between the neutralino and the stau next-to-lightest supersymmetric particle (crucial to quantify the coannihilation effect), the resulting prediction for the relic abundance spans four orders of magnitude. This result clearly illustrates how the LHC might be unable to quantify the neutralino relic abundance and thus to determine whether or not it is the main ingredient of the DM. 
Furthermore, the posterior pdf shows a multimodal structure, with two fairly well separated probability density peaks indicating two physically different solutions.

\begin{figure*}
\epsfig{file=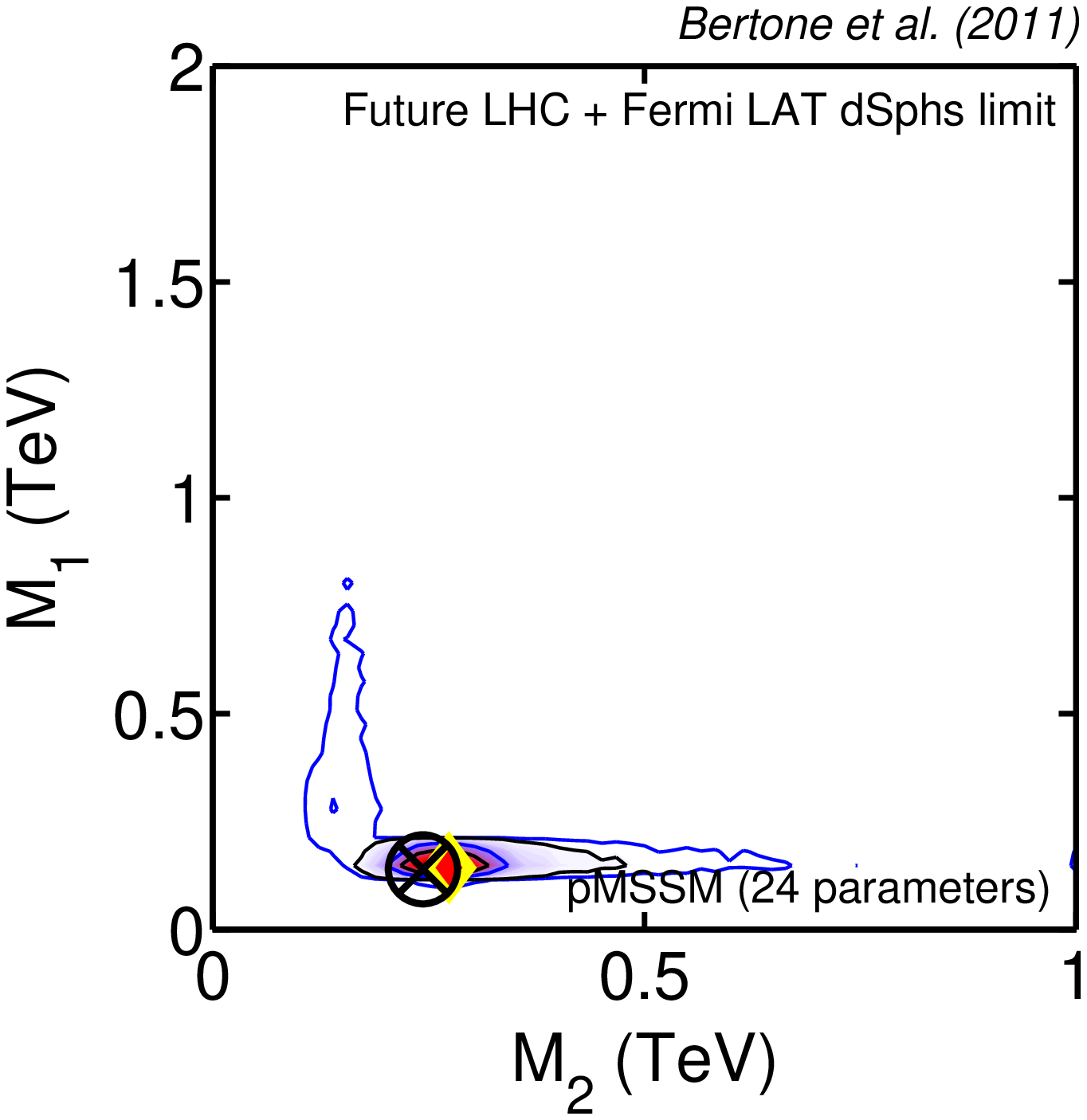,width=0.45\textwidth}
\epsfig{file=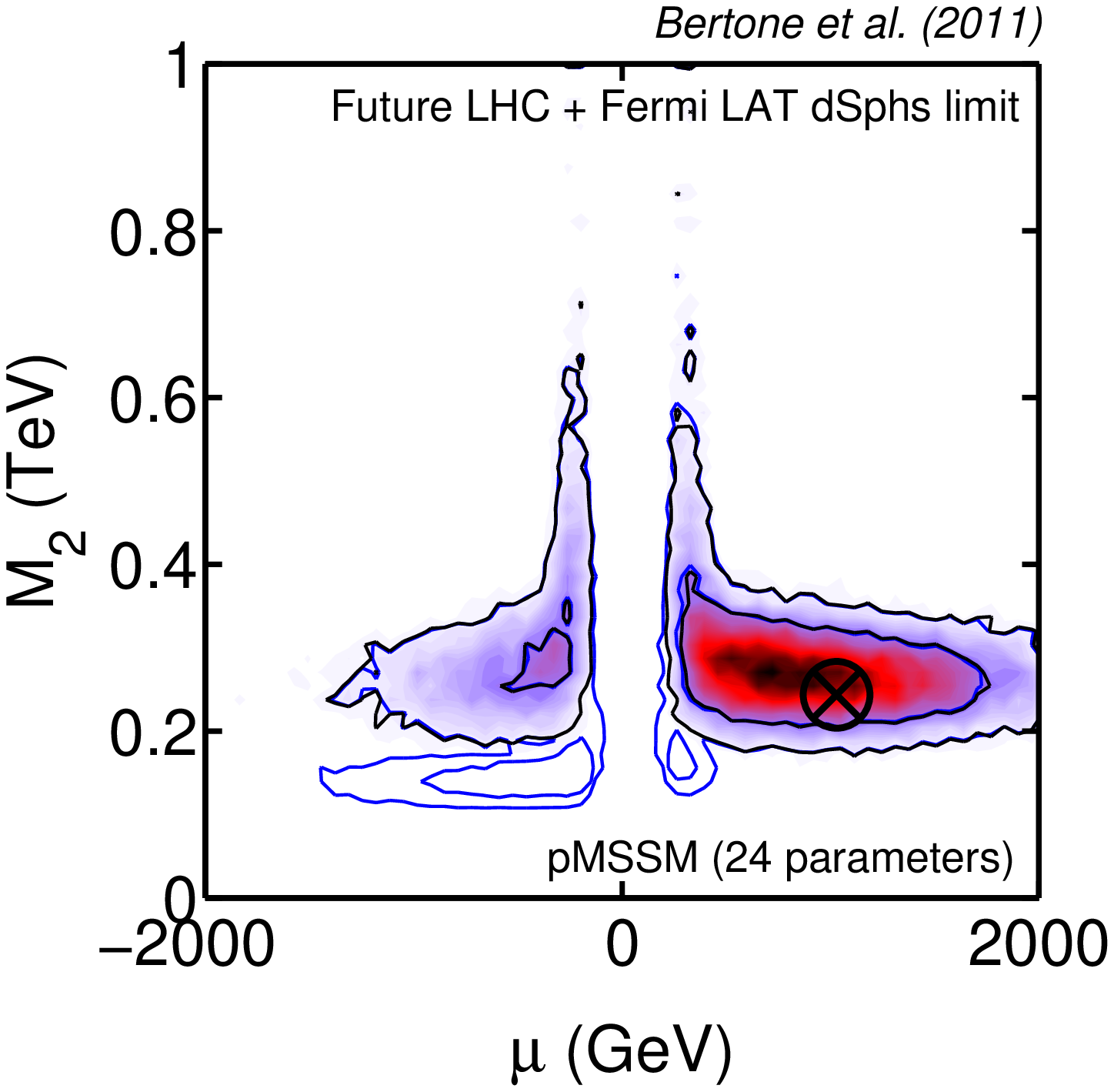,width=0.45\textwidth}
\caption{2-Dimensional marginalized posterior pdf in the $(M_1,\,M_2)$ and in the $(M_2,\,\mu)$ planes. The inner and outer contour encloses 68\% and 98\% probability regions, respectively. The empty contours are for the case where  LHC only data are applied, whereas the filled regions include current Fermi LAT upper limits from the combined analysis of dSphs. The best fit is shown with the encircled black cross while the true value is given by the yellow/red diamond.}
\label{pdfcomp}
\end{figure*}

This uncertainty in the neutralino relic density is a direct consequence of the impossibility of determining its composition in an unambiguous way \cite{Baltz:2006fm}. 
Only the two lightest neutralino mass eigenstates are measured (and none of the charginos), which is not enough to constrain the neutralino mass matrix.
We illustrate this in Fig.\,\ref{pdfcomp}, where 
the posterior pdfs for the parameters in the neutralino mass matrix ($M_1$, $M_2$, and $\mu$) obtained with LHC only data are shown by the empty contours. 
As we can observe, there are two possible solutions that satisfy the LHC constraints: one in which the neutralino is mostly Bino-like ($M_1<M_2<\mu$) and another one in which it is mostly Wino-like ($M_2<M_1<\mu$). 
Moreover, the  $\mu$ parameter is not well determined (since the heavier mass eigenstates are not measured) and varies in a wide range\footnote{Notice that although in theory we could have also obtained a solution in which the second-lightest neutralino was Higgsino-like (that is, $\mu<M_2$), this possibility is constrained by the determination of masses in the Higgs sector, since a light pseudoscalar would have also been present.}. This implies that the Higgsino composition of the lightest neutralino can vary significantly.

From the discussion above it is easy to identify which is the neutralino composition associated to the different solutions of the relic density in the pdf. The peak with $\relic h^2\gtrsim 0.1$ in Fig.\,\ref{pdfLHC} corresponds to points in the parameter space in which the neutralino is Bino-like (thus having a smaller annihilation cross section and, consequently, a larger relic abundance, compatible with the relic density measured with cosmological data). The long tail originating in this maximum and extending towards smaller values of $\relic h^2$ is obtained for neutralinos with an increasing value of the Higgsino component (this is, those for which the $\mu$ parameter is smaller in Fig. \ref{pdfcomp}). 
Finally, the second peak situated at $\relic h^2\approx 10^{-3}$ corresponds to Wino-like neutralinos, which annihilate very efficiently in the early Universe.

\section{Indirect detection constraints}  

The indirect detection signals discussed here (gamma-rays from dwarfs and modifications of CMB spectrum) depend on the neutralino self-annihilation cross section, $(\sigma v)$, on the neutralino mass, $m_{\tilde{\chi}_1^0}$ and on the spectrum of standard model particles produced in the annihilation of neutralinos, which enters in the calculation of the total photon spectrum per annihilation (relevant for the search of gamma-rays from dwarfs) and of the fraction of energy that couples with the gas during recombination (relevant for CMB constraints). 
Indirect searches can therefore be used to constrain these parameters, under specific assumptions on astrophysical quantities, and as we shall see, they allow to exclude portions of the phenomenological parameter space that would remain viable under future high-luminosity LHC measurements. 

\subsection{Fermi LAT constraints from dwarf Spheroidal galaxies}
\label{sec:gammas}

The first 24 months of data obtained by the Fermi LAT telescope in survey 
mode have been analyzed in Ref. \cite{Collaboration:2011wa} to search for 
gamma-ray emission from the position of 10 dShps including Draco.
The lack of detection allowed to set constraints on the gamma-ray emission 
from each dSph and, assuming a certain DM content, on the DM 
self-annihilation cross section $(\sigma v)$.
The gamma-ray flux at energy $E_\gamma$ due to DM annihilations from the 
direction $\Psi$ is given by: 
\begin{equation}
\frac{d \Phi_\gamma}{dE_\gamma}(E_\gamma, \Psi)=
 \frac{1}{4 \pi} \frac{(\sigma v) }{2 m_{\tilde{\chi}_1^0}^2} \,
\frac{d N_\gamma}{d E_\gamma} 
\int_{\Delta\Omega} \int \frac{\rho^2(r(s,\Psi))}{s^2} ds \, d\Omega
\label{flusso}
\end{equation}
where $d N_\gamma / dE_\gamma= \sum_{f} d N^f_\gamma / dE_\gamma$ is the total 
differential photon spectrum per annihilation, obtained by adding up the 
contributions of all annihilation channels $f$, weighted by the corresponding 
branching ratio $B_f$.
The DM distribution in the dwarf galaxies $\rho(r)$ is assumed to be 
spherically symmetric, and therefore it is a function only of the radius 
$r$, which can itself be expressed as a function of the distance along the 
line-of-sight from the observer $s$, and the angle with respect to the center 
of the dwarf $\Psi$. 
To obtain the annihilation rate, the square of the DM density is then 
integrated along the line-of-sight $s$ over the solid angle $\Delta\Omega$.

If no excess emission is detected from the direction of a dwarf (which is 
the case so far), then Eq.~\eqref{flusso} allows one to translate an upper 
limit in flux into an upper limit on the DM parameters, once a specific DM 
profile is assumed for the dSph.
In Ref. \cite{Collaboration:2011wa}, the Fermi LAT collaboration combined
the data from 10 dSphs into a single likelihood analysis, obtaining an upper
limit on $(\sigma v)$ of the order of $10^{-25}$cm$^{3}$s$^{-1}$ for a DM mass 
around 130 GeV (in the case of annihilation into $b\bar{b}$). The analysis
in Ref. \cite{Collaboration:2011wa} accounts for the astrophysical 
uncertainties on the DM profile for each dSph. The DM profile can be 
determined from kinematic data of the member stars, and in particular 
measurements of stellar velocity dispersion can be used to build a likelihood 
function that depends on the parameters defining a DM halo profile 
\cite{Martinez:2009jh}. 
These quantities have then been included in the likelihood analysis of Fermi 
LAT data, so that their final result accounts for our relatively poor 
knowledge of DM in dSphs (see Ref. \cite{Collaboration:2011wa} for more 
details). 

We focus here for definitiveness on the upper limit on $(\sigma v)$ derived 
for a DM particle annihilating to $b$ quarks, taken from Fig. 2 of 
Ref. \cite{Collaboration:2011wa}. That same figures also indicates how the 
upper limit depends on the dominant annihilation channel. With a neutralino 
mass around 130 GeV, as is the case for our benchmark point, the constraints 
are all within a factor of two from the case of annihilation into $b$ quarks, 
and this is not enough to change our results significantly (see later). The 
only exception is for an annihilation predominantly into $\mu^+\mu^-$, for 
which the Fermi LAT upper limit is approximately an order of magnitude 
weaker. However, we checked that the branching ratio into muons is 
subdominant (with a branching ratio smaller than about 0.1) for all the samples 
in our scan, once LHC data are included, hence this case can be discounted. 

\begin{figure*}
\epsfig{file=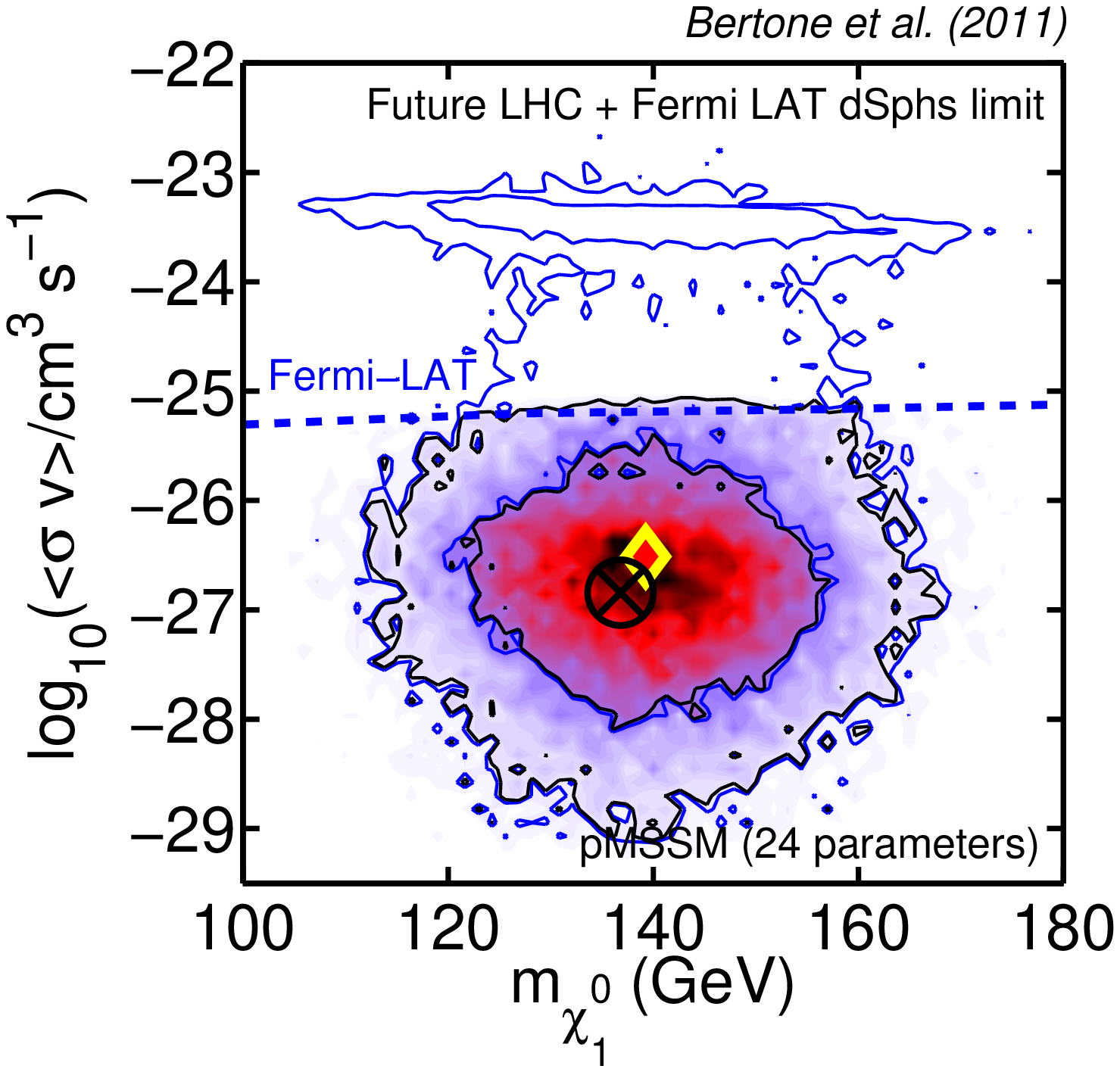,width=0.3\textwidth}
\epsfig{file=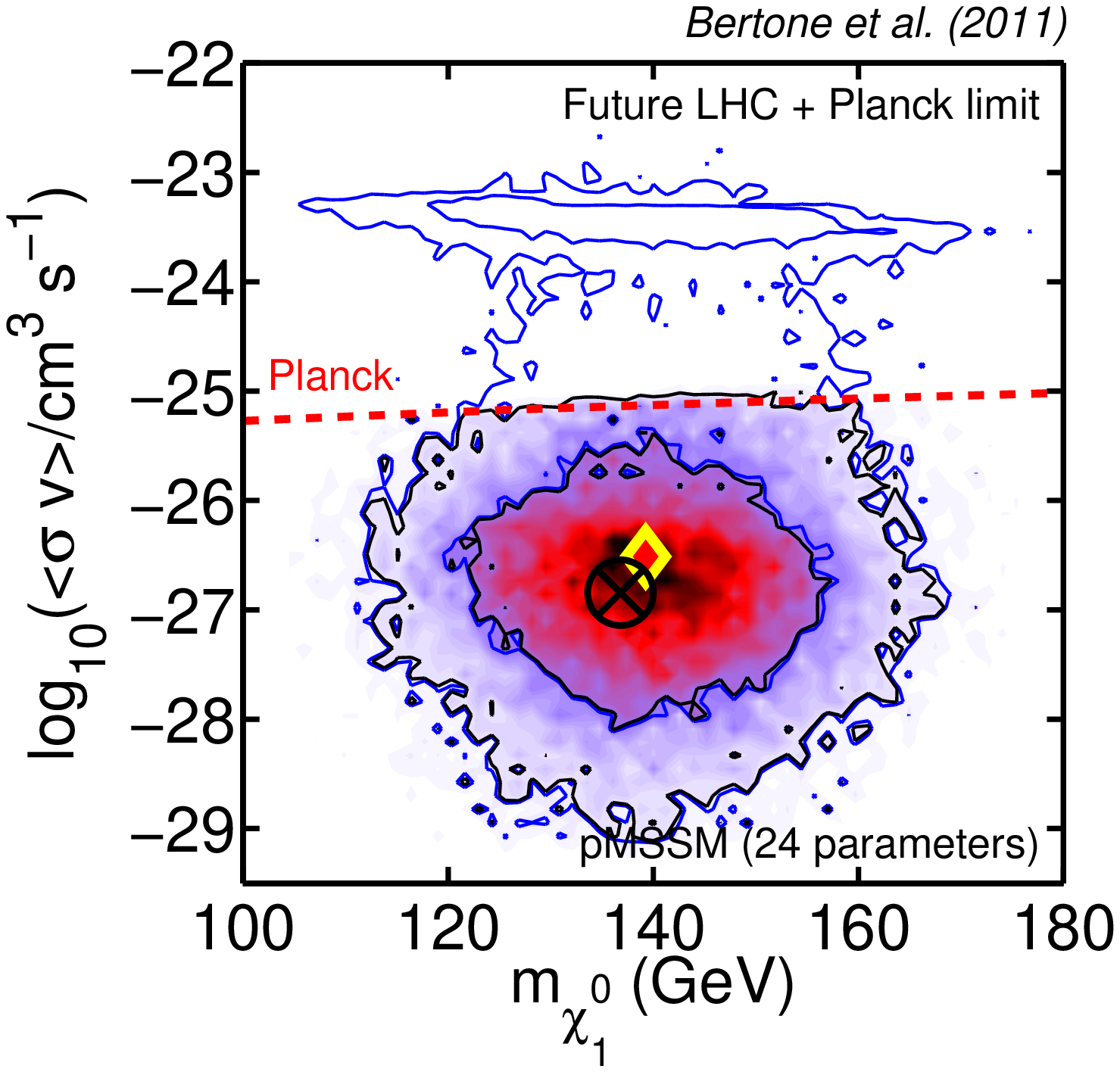,width=0.3\textwidth}
\epsfig{file=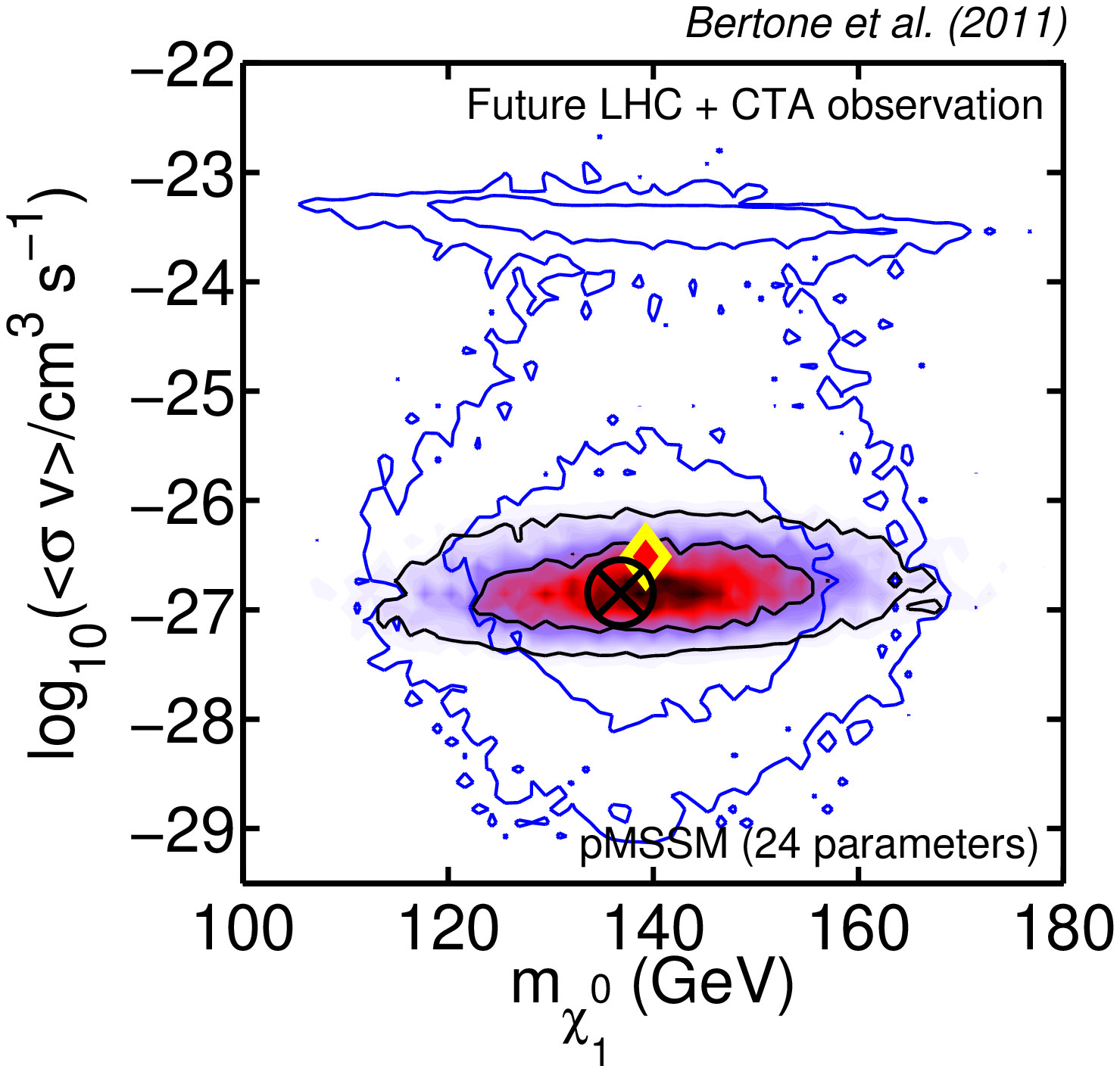,width=0.3\textwidth}
\newline
\epsfig{file=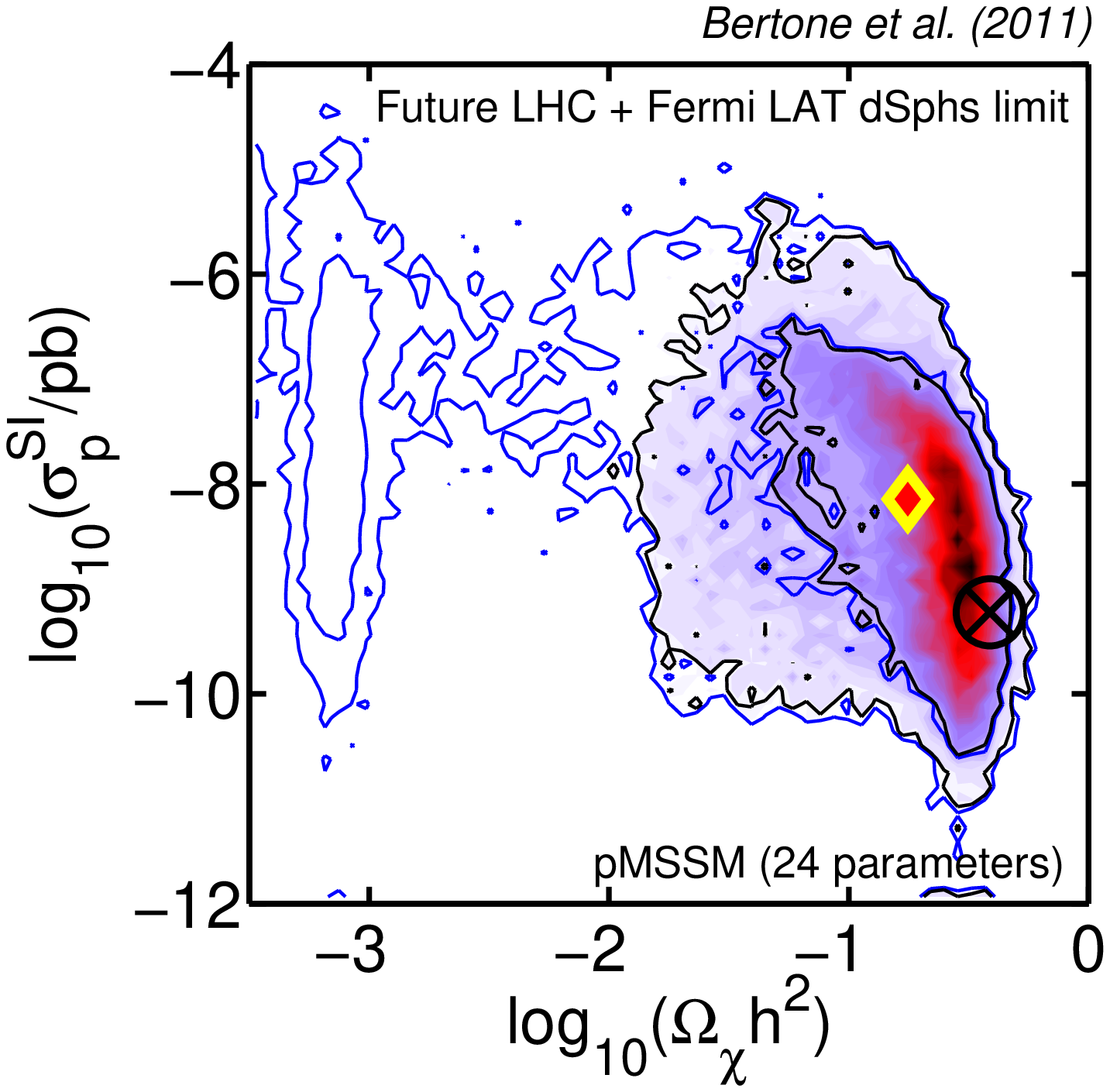,width=0.3\textwidth}
\epsfig{file=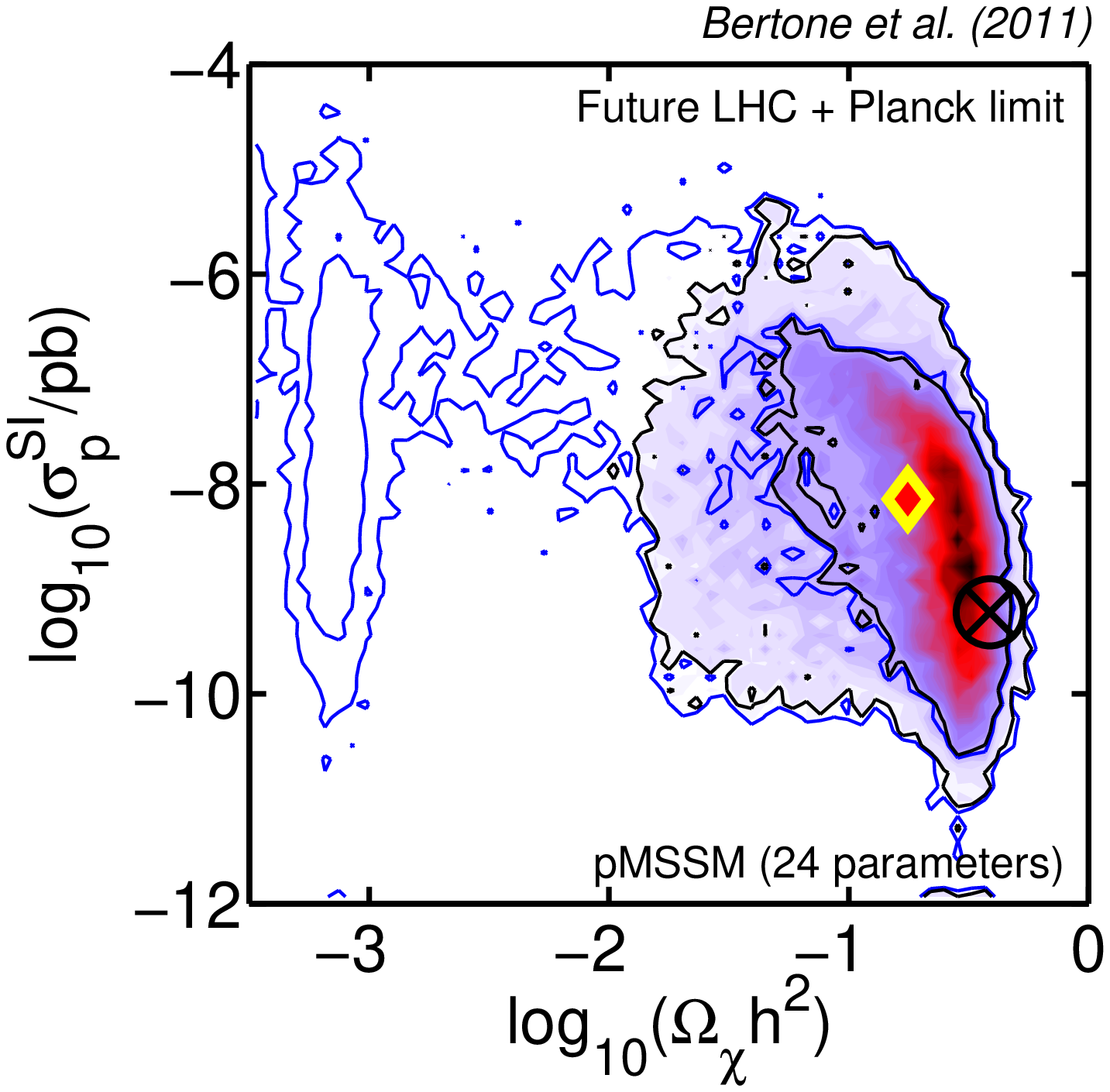,width=0.3\textwidth}
\epsfig{file=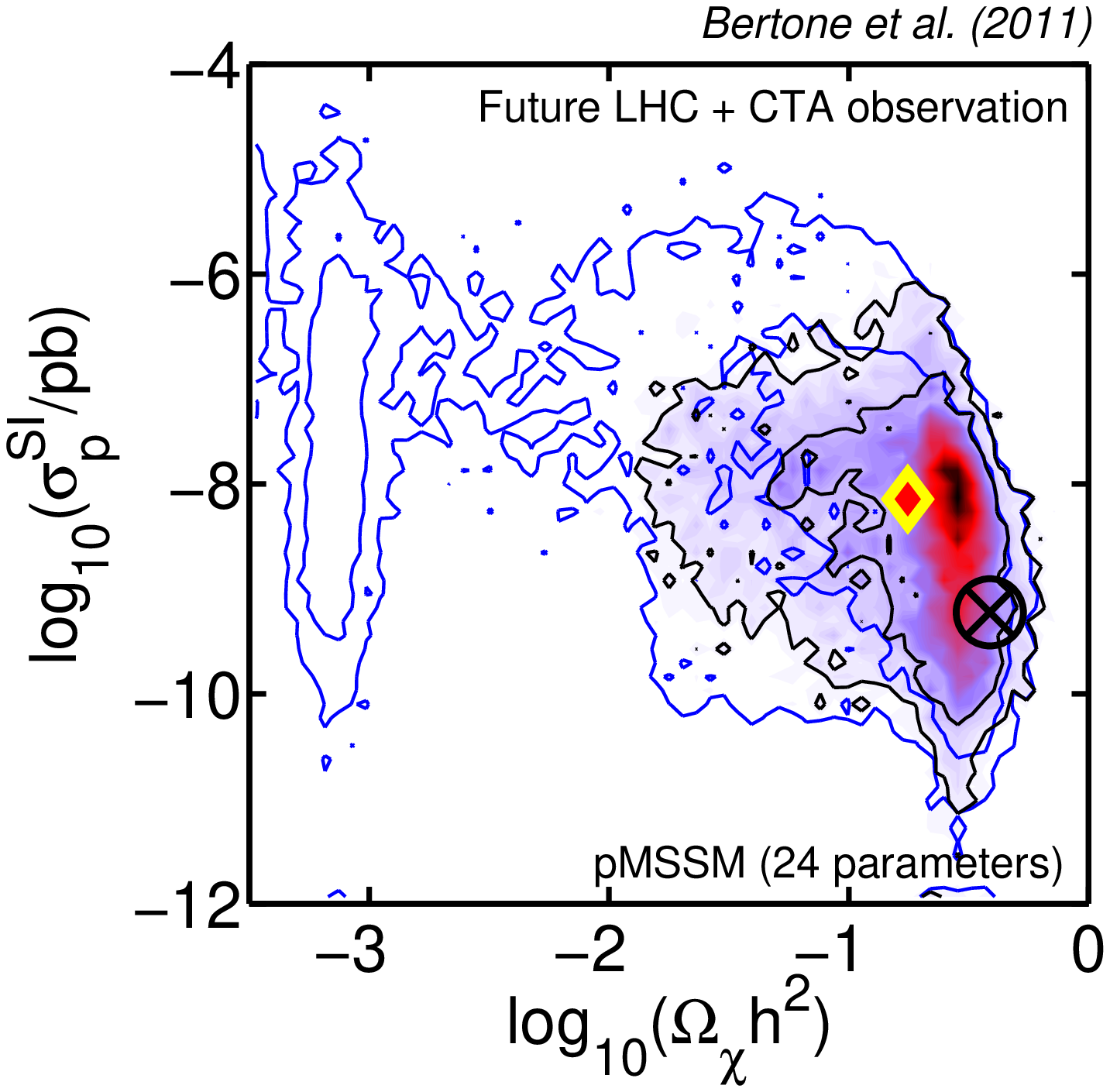,width=0.3\textwidth}
\newline
\epsfig{file=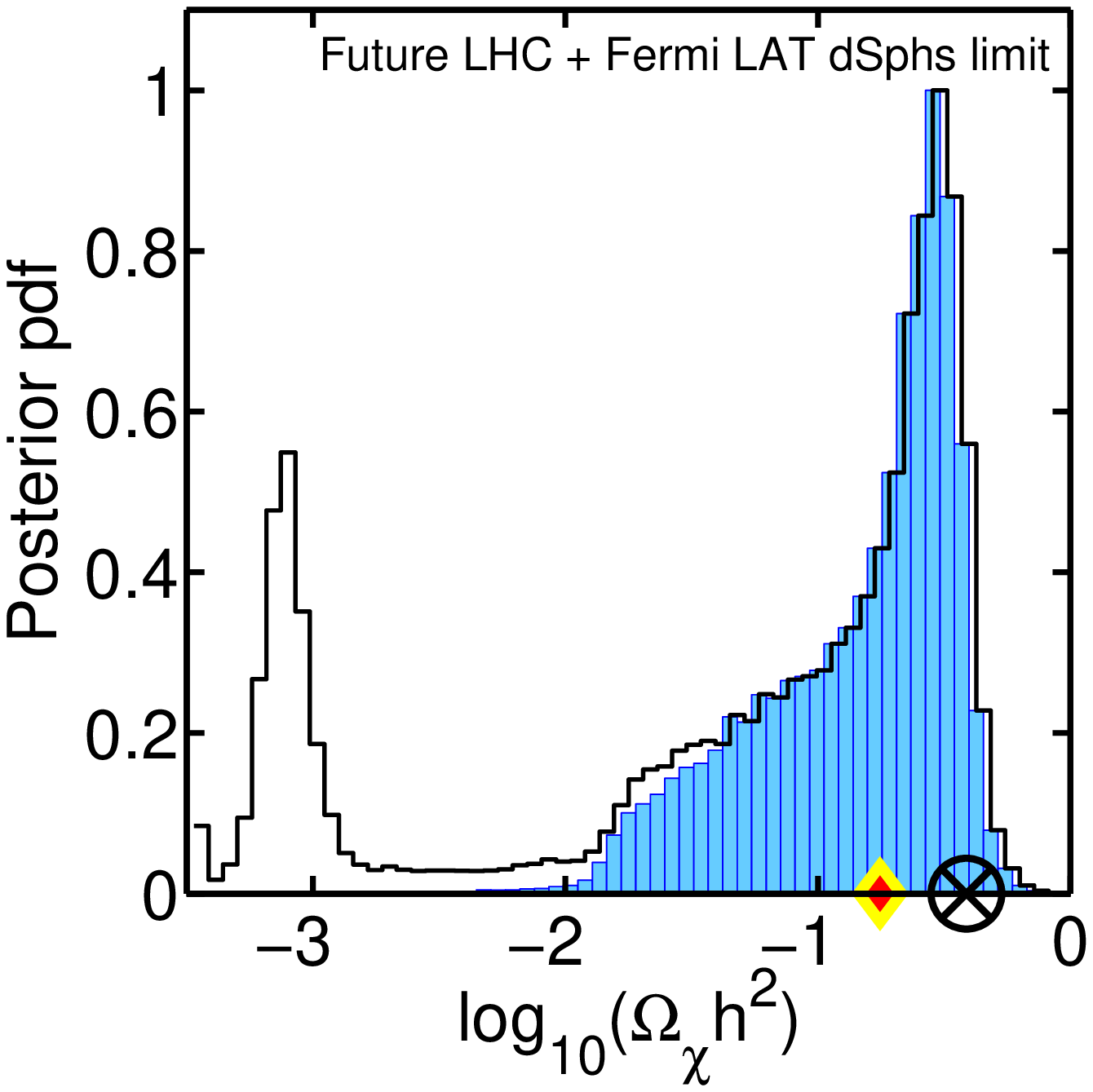,width=0.3\textwidth}
\epsfig{file=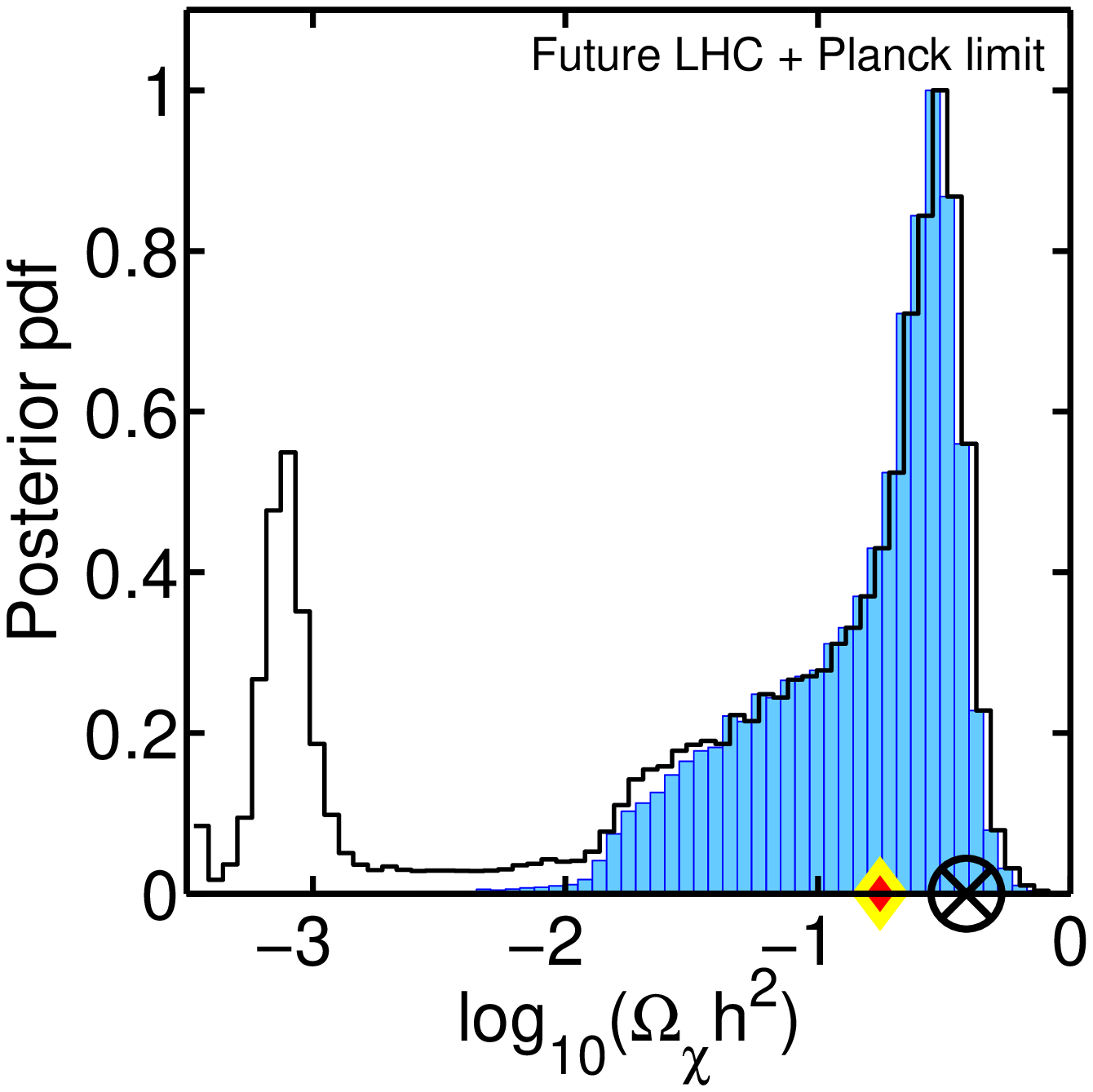,width=0.3\textwidth}
\epsfig{file=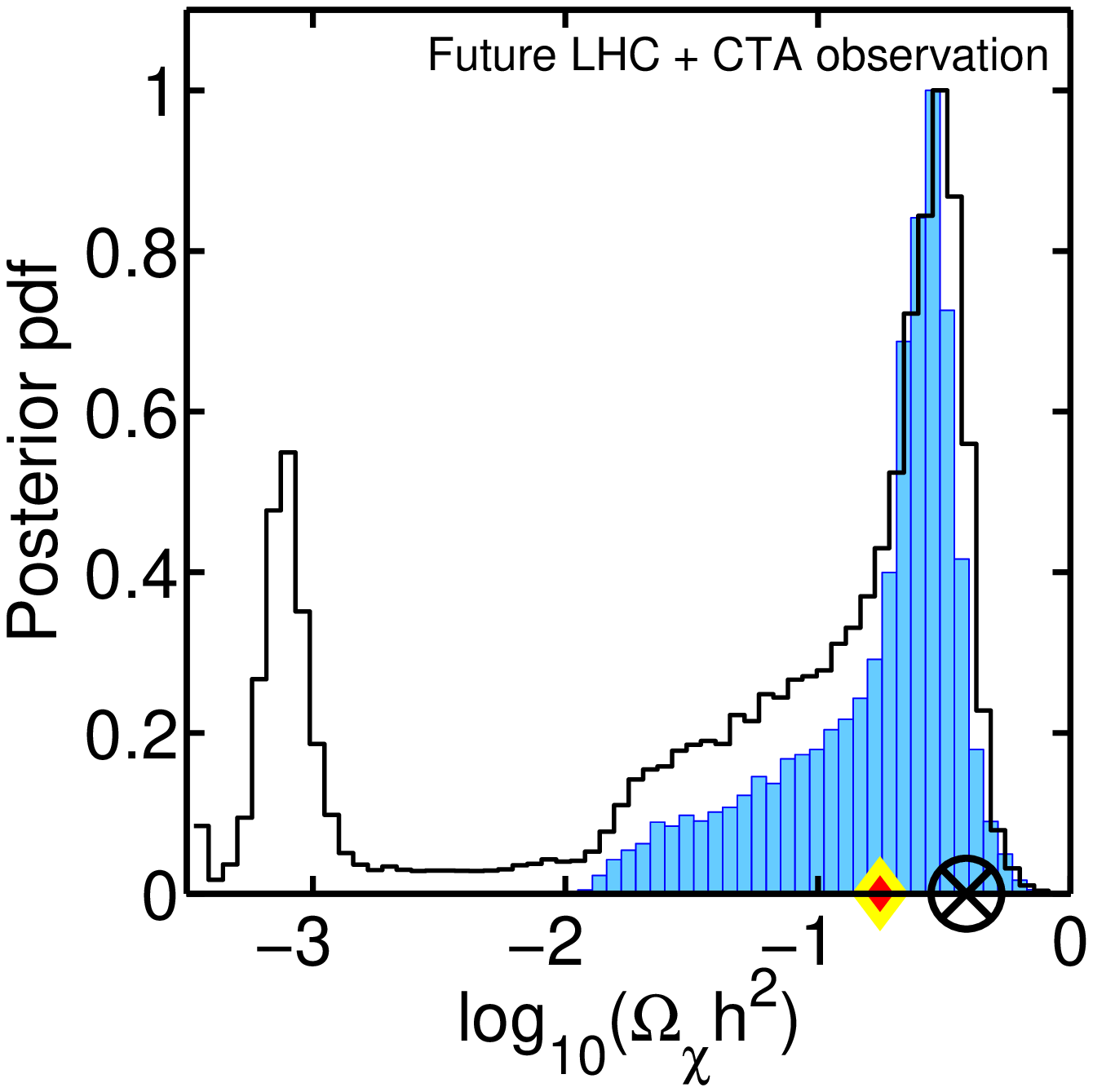,width=0.3\textwidth}
\caption{\label{fig:plots} 2-Dimensional marginal posterior pdf in the plane $(m_\chi,(\sigma v))$ (top row) and $(\Omega_\chi h^2,\sigma_{\chi-p}^{\mbox{\tiny{SI}}}$) (middle row), including simulated future LHC data only (blue/empty contours) and adding Fermi LAT upper limit from the combined analysis of dSphs (left column, filled regions). The second columns combines LHC data with the upper limit expected from Planck on the reionization of the CMB radiation, while the third column combines LHC data with an hypothetical detection of gamma-rays from the Draco dSph obtained with CTA. The bottom row shows the 1D marginal pdf for the relic density $\Omega_\chi h^2$. The inner and outer contour encloses 68\% and 95\% probability regions, respectively. The best fit is shown with the encircled black cross while the true value is given by the yellow/red diamond. The black continous line in the bottom row indicates the pdf of $\Omega_\chi h^2$ for the LHC-only case.}
\end{figure*}

We include the information provided by Fermi LAT on the combined analysis
of the 10 dSphs by assigning a likelihood of 0 to all samples in our LHC-only scans that
have an annihilation cross section larger than the 95\% upper limit in Fig.~2
of Ref.~\cite{Collaboration:2011wa}. A more detailed analysis would include the full likelihood function in a more refined way, but this is not necessary for the purpose of our study. We stress that in order to implement the Fermi LAT constraints we make the 
assumption that the neutralino found at the LHC contributes 100\% to the 
non-baryonic DM in dwarf galaxies, similarly to the 
{\it consistency check} approach of Paper I.
The resulting two-dimensional posterior pdf in the planes 
$(m_\chi, \sigma v )$ and $(\Omega_\chi h^2,\sigma_{\chi-p}^{\mbox{\tiny{SI}}}$) are 
shown in the first column of Fig.~\ref{fig:plots} (filled contours), where 
they are compared with the case where LHC only future contraints are used 
(blue/emtpy contours). Focusing first on the top panel, we notice that the LHC alone (empty contours) is not going to be able to identify the correct mode in the posterior distribution. The secondary mode, at large values of $\sigma v$, corresponds to the case where the neutralino is Wino-like. This solution can however be ruled out once the Fermi LAT information from dSphs is included in the likelihood (filled contours). The resulting identification of the correct cosmological solution in terms of the predicted relic density is confirmed by the plot in the second panel, showing how Fermi LAT limits can eliminate the mode in the distribution corresponding to subdominant relic density. 
The last row of Fig.~\ref{fig:plots} shows the 1D 
marginalized pdf for the relic density, comparing the LHC only constraints (empty histogram), given in Fig.~\ref{pdfLHC}, with what can be otbained by combining LHC with Fermi LAT. We see clearly that the Wino-like neutralino solution disappears in this latter case. 

The impact of Fermi LAT data can also 
be observed in the reconstruction of the neutralino mass parameters of 
Fig.\,\ref{pdfcomp}, where Fermi can rule out the 
region where $M_2<M_1$. 
However, Fermi has little impact in the reconstruction of the 
$\mu$ parameter, and therefore the Higgsino composition of the neutralino 
is not well determined. As a consequence, the reconstruction of the 
neutralino relic density (bottom panel of Fig.~\ref{fig:plots}) still displays the characteristic long tail towards
 small values of $\relic h^2$. In practice, even the combined data can only constrain the relic density within an order of magnitude or so of the true value.

\subsection{Planck constraints from recombination history}

The most robust constraints on the annihilation cross section are perhaps 
those arising from observations of the Cosmic Microwave Background (CMB) 
radiation \cite{Galli:2009zc,Slatyer:2009yq,Hutsi:2011vx,Finkbeiner:2011dx,Galli:2011rz}, since they do not depend on 
poorly known quantities such as the DM profile at the Galactic center or in dwarf galaxies. 

The annihilation of DM particles around redshift 
$\sim 1000$ inevitably affects the processes of recombination and 
reionization, modifying 
the evolution of the free electron fraction $x_e$ and the temperature of 
baryons.
The evolution of the ionization fraction $x_e$ satisfies the following
equation:
\begin{equation}
\label{eq:dxe}
\frac{d x_e}{d z} = \frac{1}{(1+z)H(z)}
\left[ R_s(z) - I_s(z) - I_X(z)  \right],
\end{equation}
where $R_s$ is the standard recombination rate 
\cite{Peebles:1968ja,Zeldovich:1969en}, $I_s$ the ionization rate 
by standard sources and the $I_X$ term represents a ``non--standard'' source 
of ionization; in our case it takes into account that, during recombination, 
annihilations of DM particles increase the ionization rate both by direct 
ionization from the ground state, and by contributing additional 
Lyman-$\alpha$ photons.
Therefore, the ionization rate due to DM annihilations can be written as:
\begin{equation}
I_X(z) = I_{Xi}(z) + I_{X\alpha}(z),
\label{terms}
\end{equation}
where $I_{Xi}$ is the ionization rate due to ionizing photons, and $I_{X\alpha}$
the ionization rate due to additional Lyman-$\alpha$ photons.
The rate of energy release $dE(z)/dt$ per unit of volume by a relic 
self-annihilating DM particle at redshift $z$ is given by
\begin{equation}
\label{enrateselfDM}
\frac{dE}{dt}(z)= 
\rho^2_c c^2 \Omega^2_{DM} (1+z)^6 f  \frac{\sigma v}{m_\chi}.
\end{equation}
Here, $\Omega_{DM}$ is the DM density parameter and $\rho_c$ the critical density 
of the Universe today; the parameter $f$ indicates the fraction of energy 
which is absorbed {\it overall} by the gas, under the approximation that 
the energy absorption takes place locally. The latter parameter, together
with the DM mass and annihilation cross section, define the parameter we call $p_\text{ann} \equiv f  \sigma v / m_\chi $.

Both the non-standard ionization rates $I_{Xi}$ and $I_{X\alpha}$ are related 
to the energy release rate as follows:
\begin{equation}
I_{Xi} = \frac{C \, \chi_i}{n_H(z) E_i} \frac{dE}{dt}(z) \, , \, I_{X\alpha} = \frac{(1-C) \; \chi_\alpha}{n_H(z) E_\alpha} \frac{dE}{dt}(z) 
\end{equation}
where $E_i$ is the average ionization energy per baryon, $E_\alpha$ is the 
difference in binding energy between the $1s$ and $2p$ energy levels of a 
hydrogen atom, $n_H(z)$ is the number density of hydrogen nuclei and 
$\chi_i=\chi_\alpha=(1-x_e)/3$ are the fractions of energy going to 
ionization and to Lyman-$\alpha$ photons, respectively. 
A fraction of the energy released by annihilating DM particles goes 
into heating of baryonic gas, adding an extra term in the standard 
equation for the evolution of the matter temperature (see Eqs.~(9) and (10) of 
Ref. \cite{Galli:2009zc}).

In Ref.~\cite{Galli:2009zc} the angular 
power spectrum of CMB anisotropies was computed, taking into account the presence of DM 
annihilations and  an upper limit of 
$2.0 \times 10^{-6}$m$^3$s$^{-1}$kg$^{-1}$ was derived from WMAP5 data at the 95\% 
confidence level. This limit was already sufficient to exclude models 
that exhibit a large Sommerfeld enhancement for $(\sigma v)$, such as those 
proposed to fit the PAMELA and ATIC results. 
The upper 
limit is expected to become an order of magnitude more stringent, of order $1.5 \times 10^{-7}$ m$^3$s$^{-1}$kg$^{-1}$, with upcoming 
Planck data. In order to study the impact of such data on LHC constraints on the pMSSM parameters, we have added a projected upper limit on $p_\text{ann}$ from Planck, shown as the dashed red line in the top central panel of Fig.~\ref{fig:plots}.  
%
In doing so, we adopted a fixed value $f=0.5$ for the fraction of energy absorbed by the gas. Although the actual value of $f$ depends on the particle physics characteristics of the neutralino, detailed estimations of $f$ (e.g., Ref. \cite{Slatyer:2009yq}) show that this is a reasonable choice.

As shown in the central panels of Fig. \ref{fig:plots}, we find that future Planck constraints will complement LHC data in a very similar way to those arising from the non-detection of gamma-rays from Fermi LAT. The combination of LHC and Planck will exclude regions of the parameter space with high annihilation cross section and low relic density, leading again to the indentification of the correct cosmological solution for the relic density within an order of magnitude (bottom panel).

\subsection{Constraints from a CTA-like experiment}

We now move on to analyze the implications that an actual detection of excess photons would have on the parameter space of SUSY DM discovered at the LHC. Cherenkov telescopes represent the future (or at least the near future) of gamma-ray experiments. Cherenkov telescopes detect gamma-rays indirectly, through the detection
of the electromagnetic shower produced by the interaction of primary gamma-rays with
the atmosphere. The shape of the image created by the shower in the telescope
camera allows to discriminate photons from hadrons, and it also provides information
on the incident gamma-ray as, e.g. energy and direction (see 
Ref. \cite{Brun:2011rg} for a recent review on the Cherenkov telescopes 
and their role in DM searches).
We focus here on the upcoming Cherenkov Telescope Array (CTA)
\cite{Consortium:2010bc}. Current plans are to build a facility on two sites, one on each hemisphere, with 
 telescopes of three different sizes, for a total of about 100 instruments. Such 
large number of telescopes, combined with a large field of 
view, will allow one to monitor more sources at once, and also to perform  
surveys of large portions of the sky \cite{Brun:2010ci,Glicenstein}.
CTA sensitivity is estimated to go down by a factor of 10 with respect
to current Cherenkov telescopes, and the energy range will cover the interval from 10 GeV to 200 TeV 
\cite{Consortium:2010bc,Bernloehr:2008xd}.
The CTA collaboration will complete its Design Phase in two years from now. 
Currently different prototypes of the final telescopes are already under 
construction and will be tested. A realistic timescale for the construction of the experiment is 5 years from now.

For our purposes, we will consider a rather optimistic energy threshold of 
20 GeV. Our benchmark point corresponds to a neutralino mass of 139.3 GeV
(see Tab. \ref{tab:constraints}), so gamma-rays from DM annihilation are
expected to fall in an energy range for which a reasonable estimation of
CTA effective are is $10^{4}$m$^2$ (contrary to the largest values predicted
for higher energies).

We consider CTA observations of Draco, and we assume for its astrophysical 
factor an informative prior $p(\lambda)$, modeled as a Gaussian distribution with mean $1.20 \times 10^{19}$ GeV cm$^{-5}$sr and standard deviation of
$0.31 \times 10^{19}$GeV cm$^{-5}$sr, following Tab.~4 of 
Ref.~\cite{Abdo:2010ex}.
With such prescriptions (and for an astrophysical factor equal to the mean
of the Gaussian prior), our benchmark SUSY point predicts a flux of
 $N_{\mbox{\tiny{CTA}}}=7.1$ photons from the halo of Draco above 20 GeV, with 
1000 hours of observation. As we are interested in the constraining power of CTA around the true value of the parameters, we neglect Poisson noise in realizing the counts and we thus assume an observed number of photons above 20 GeV  $\hat{N}_{\mbox{\tiny{CTA}}}=7$. There is not enough statistics to build a gamma-ray 
spectrum, so we will only consider the information coming from the detection 
of all photons above the energy threshold. This is implemented in the 
likelihood as an additional experimental constraint of the following form:
\begin{equation}
\like_\text{CTA}({\mathbf x}) =  
\int p(\lambda N({\mathbf x})  | \hat{N}_\text{CTA}) p(\lambda) d\lambda
\end{equation}
where $N({\mathbf x})$ is the prediceted number of photons above threshold for a pMSSM parameters value $\mathbf x$ and an astrophysical factor equal to unity, $\lambda$ is the unknown true astrophysical factor for Draco (which we marginalize over), $ p(\lambda N({\mathbf x}) | \hat{N}_\text{CTA}) $ is a Poisson distribution in $\hat{N}_\text{CTA}$ and $p(\lambda)$ represents its prior.

The results of augmenting LHC data with a future CTA-like detection can be seen in the last column of Fig.~\ref{fig:plots}. With a detection of only 7 photons from Draco, the LHC constraints on the DM self-annihilation cross section collapse from 6 orders of magnitude down to about 1 order of magnitude. As a consequence, the relic abundance can again be constrained around its true value. We point out that this dramatic improvements hold true even when, like we do here, astrophysical uncertainties are marginalized over. 

\section{Conclusions}

We have discussed the complementarity of indirect and accelerator DM searches, and we have shown that current upper limits on the DM self-annihilation cross section arising from Fermi LAT constraints on gamma-ray from dwarfs are already sufficiently powerful to break degeneracies in the phenomenological parameter space of DM that would be present even in high-luminosity LHC data. Similar results are obtained if we consider the information that Planck will provide on the distortion of the CMB, or if we assume a possible future detection of gamma-rays from Draco by the CTA telescope.

In Paper I, we had performed a similar analysis for a combination of accelerator and direct detection data, and demonstrated how a simple yet powerful {\em scaling Ansatz} (stating that the local contribution of neutralinos to Galactic DM was the same as on average in the Universe) led to the correct identification of the cosmological solution. In this paper, we focused on the power of indirect detection techniques, which do not require such an Ansatz. Instead, we assumed that the neutralino seen in colliders makes up the totality of the DM in the Universe (this is what in Paper I was called ``the consistency check'' approach).  This corresponds to asking the question: can the observed neutralinos make all the DM in the local universe? This approach allows one to discard solutions corresponding to high annihilation cross sections without making any assumption on the expansion rate of the Universe at freeze-out, therefore bypassing all the difficulties arising from all the effects that might possibly modify it (see Ref. \cite{Gelmini:2010zh} for a recent review).

The removal of the spurious Wino-like mode which would survive with LHC measurements alone and the identification of the correct cosmological solution can be achieved via various indirect detection channels: current Fermi LAT limits on the flux from dwarfs are already sufficient to this goal (at least for the benchmark point studied here), but Planck constraints will give similar results when they become available in 2013, while probing very different physics. Furthermore, we have shown that a detection of gamma-ray from Draco from CTA can lead to similarly stringent conclusions, even when astrophysical uncertainties are included in the analysis.

This work therefore makes the case for a vigorous program of indirect searches, especially in the case where the LHC finds evidence for New Physics. We have demonstrated that in this case indirect detection methods can have a crucial role to play, {\it even in the case where the experiments only return upper limits to the signal}. Together with the complementarity between LHC and direct detection methods presented in Paper I, our results thus highlight the fundamental importance of a multi-pronged approach to DM identification.

{\em Acknowledgements} R.T. would like to thank the University of Zurich for hospitality. M.F. and D.G.C would like to thank the Spanish MICINN's Consolider-Ingenio 2010 Programme under grant MultiDark CSD2009-00064. D.G.C. is supported by the Ram\'on y Cajal program of the Spanish MICINN and also acknowledges support from the MICINN under grant FPA2009-08958, the Community of Madrid under grant HEPHACOS S2009/ESP-1473, and the European Union under the Marie Curie-ITN program PITN-GA-2009-237920. The work of G.B. is supported by the ERC Starting Grant {\it WIMPs Kairos}.


\begin{thebibliography}{99}
\bibitem{Bertone:2010rv}
  G.~Bertone, D.~G.~Cerdeno, M.~Fornasa, R.~R.~de Austri, R.~Trotta,
  Phys.\ Rev.\  {\bf D82}, 055008 (2010).
  
\bibitem{Jungman:1995df}
  G.~Jungman, M.~Kamionkowski and K.~Griest, 
  {\it Phys.\ Rept.\ } {\bf 267} (1996) 195.

\bibitem{Munoz:2003gx}
  C. Mu\~noz,
  {\it Int. J. Mod. Phys.} {\bf A19} (2004) 2093. 

\bibitem{Bertone:2004pz}
  G.~Bertone, D.~Hooper and J.~Silk,
  Phys.\ Rept.\  {\bf 405} (2005) 279. 

\bibitem{book}
  {\it Particle Dark Matter: Observations, Models and Searches}, ed. G. Bertone, 2010, Cambridge University Press.

\bibitem{Bertone:2010at}
  G.~Bertone,
  Nature {\bf 468}, 389-393 (2010).
  
\bibitem{Baltz:2006fm}
  E.~A.~Baltz, M.~Battaglia, M.~E.~Peskin and T.~Wizansky,
  Phys.\ Rev.\  D {\bf 74} (2006) 103521.

\bibitem{Green:2007rb}
  A.~M.~Green,
  JCAP {\bf 0708} (2007) 022.
  
\bibitem{Green:2008rd}
  A.~M.~Green,
  JCAP {\bf 0807} (2008) 005.

\bibitem{Drees:2008bv}
  M.~Drees and C.~L.~Shan,
  JCAP {\bf 0806} (2008) 012.

\bibitem{Bertone:2007xj}
  G.~Bertone, D.~G.~Cerde\~no, J.~I.~Collar and B.~C.~Odom,
  Phys.\ Rev.\ Lett.\  {\bf 99} (2007) 151301.

\bibitem{Fermi}
  http://fermi.gsfc.nasa.gov

\bibitem{Atwood:2009ez}
  W.~B.~Atwood {\it et al.}  [LAT Collaboration],
  Astrophys.\ J.\  {\bf 697} (2009) 1071.

\bibitem{HESS}
  http://www.mpi-hd.mpg.de/hfm/HESS/pages/about/te\-lescopes/

\bibitem{Aharonian:2005zz}
  F.~A.~Aharonian, H.~J.~Volk and D.~Horns,
  {\it  AIP Conf. Proc. 745 (2005) 802 p}.

\bibitem{IceCube}
  http://icecube.wisc.edu/

\bibitem{Halzen:2010yj}
  F.~Halzen and S.~R.~Klein,
  Rev.\ Sci.\ Instrum.\  {\bf 81} (2010) 081101.

\bibitem{Hooper:2010mq}
  D.~Hooper, L.~Goodenough,
  [arXiv:1010.2752 [hep-ph]].

\bibitem{Jeltema:2008hf}
  T.~E.~Jeltema, S.~Profumo,
  JCAP {\bf 0811 } (2008)  003.
  
\bibitem{Regis:2008ij}
  M.~Regis, P.~Ullio,
  Phys.\ Rev.\  {\bf D78 } (2008)  043505.
  
\bibitem{Bertone:2002je}
  G.~Bertone, G.~Sigl, J.~Silk,
  Mon.\ Not.\ Roy.\ Astron.\ Soc.\  {\bf 337 } (2002)  98.
  
\bibitem{Gondolo:1999ef}
  P.~Gondolo, J.~Silk,
  Phys.\ Rev.\ Lett.\  {\bf 83 } (1999)  1719-1722.
  
\bibitem{Bergstrom:1997fj}
  L.~Bergstrom, P.~Ullio, J.~H.~Buckley,
  Astropart.\ Phys.\  {\bf 9 } (1998)  137-162.
  
\bibitem{Stoehr:2003hf}
  F.~Stoehr, S.~D.~M.~White, V.~Springel, G.~Tormen, N.~Yoshida,
  Mon.\ Not.\ Roy.\ Astron.\ Soc.\  {\bf 345 } (2003)  1313.

\bibitem{Strigari:2006rd}
  L.~E.~Strigari, S.~M.~Koushiappas, J.~S.~Bullock, M.~Kaplinghat,
  Phys.\ Rev.\  {\bf D75 } (2007)  083526.
  
\bibitem{Pieri:2007ir}
  L.~Pieri, G.~Bertone, E.~Branchini,
  Mon.\ Not.\ Roy.\ Astron.\ Soc.\  {\bf 384 } (2008)  1627.

\bibitem{Pieri:2008MNRAS}
  C.~Giocoli, L.~Pieri, G.~Tormen, 
  Mon.\ Not.\ Roy.\ Astron.\ Soc.\  {\bf 387 } (2008)  689.


\bibitem{Pieri:2009AA}    
  L.~Pieri, A.~Pizzella, E.~M.~Corsini, E.~Dalla~Bont\'a, F.~Bertola, 
  Astronomy \ \& \ Astrophysics \ {\bf 496 } (2009)  351.

\bibitem{Pieri:2009MNRAS}
  L.~Pieri, M.~Lattanzi, J.~Silk, 
  Mon.\ Not.\ Roy.\ Astron.\ Soc.\  {\bf 399 } (2009)  2033.

    
\bibitem{Baltz:2008wd}
  E.~A.~Baltz, B.~Berenji, G.~Bertone, L.~Bergstrom, E.~Bloom, T.~Bringmann, J.~Chiang, J.~Cohen-Tanugi {\it et al.},
  JCAP {\bf 0807 } (2008)  013.
    
\bibitem{Martinez:2009jh}
  G.~D.~Martinez, J.~S.~Bullock, M.~Kaplinghat, L.~E.~Strigari, R.~Trotta,
  JCAP {\bf 0906 } (2009)  014.
 
\bibitem{Zhao:2005zr}
  H.~-S.~Zhao, J.~Silk,
  Phys.\ Rev.\ Lett.\  {\bf 95 } (2005)  011301.
  
\bibitem{Bertone:2005xz}
  G.~Bertone, A.~R.~Zentner, J.~Silk,
  Phys.\ Rev.\  {\bf D72 } (2005)  103517.

\bibitem{Bertone:2006nq}
  G.~Bertone,
  Phys.\ Rev.\  {\bf D73 } (2006)  103519.
  
\bibitem{Fornasa:2007nr}
  M.~Fornasa, G.~Bertone,
  Int.\ J.\ Mod.\ Phys.\  {\bf D17 } (2008)  1125-1157.
  
\bibitem{Aharonian:2008wt}
  F.~Aharonian {\it et al.} [ HESS Collaboration ],
  Phys.\ Rev.\  {\bf D78 } (2008)  072008.
  
\bibitem{Taoso:2008qz}
  M.~Taoso, S.~'i.~Ando, G.~Bertone, S.~Profumo,
  Phys.\ Rev.\  {\bf D79 } (2009)  043521.

\bibitem{Bertone:2009kj}
  G.~Bertone, M.~Fornasa, M.~Taoso, A.~R.~Zentner,
  New J.\ Phys.\  {\bf 11 } (2009)  105016.
  
\bibitem{Sandick:2011rp}
  P.~Sandick, S.~Watson,
  [arXiv:1102.2897 [astro-ph.CO]].
  
\bibitem{Merritt:2006mt}
  D.~Merritt, S.~Harfst, G.~Bertone,
  Phys.\ Rev.\  {\bf D75 } (2007)  043517.

\bibitem{Bertone:2005xv}
  G.~Bertone, D.~Merritt,
  Mod.\ Phys.\ Lett.\  {\bf A20 } (2005)  1021.

\bibitem{Bertone:2005hw}
  G.~Bertone, D.~Merritt,
  Phys.\ Rev.\  {\bf D72 } (2005)  103502.
  
\bibitem{Pieri:2009je}
  L.~Pieri, J.~Lavalle, G.~Bertone, E.~Branchini,
  Phys.\ Rev.\  {\bf D83 } (2011)  023518.
  
\bibitem{Pieri:2010PRD}
  R.~Catena, N.~Fornengo,~M. Pato, L.~Pieri, A.~Masiero, 
  Phys.\ Rev.\  {\bf D81 } (2010)  123522.

  
\bibitem{Hooper:2011ti}
  D.~Hooper, T.~Linden,
  [arXiv:1110.0006 [astro-ph.HE]].

\bibitem{Collaboration:2011wa}
  T.~F.~-L.~collaboration,
  [arXiv:1108.3546 [astro-ph.HE]].
  
\bibitem{Galli:2009zc}
  S.~Galli, F.~Iocco, G.~Bertone, A.~Melchiorri,
  Phys.\ Rev.\  {\bf D80 } (2009)  023505.
  
\bibitem{Slatyer:2009yq}
  T.~R.~Slatyer, N.~Padmanabhan, D.~P.~Finkbeiner,
  Phys.\ Rev.\  {\bf D80 } (2009)  043526.
  
\bibitem{Hutsi:2011vx}
  G.~Hutsi, J.~Chluba, A.~Hektor, M.~Raidal,
  [arXiv:1103.2766 [astro-ph.CO]].
  
\bibitem{Finkbeiner:2011dx}
  D.~P.~Finkbeiner, S.~Galli, T.~Lin, T.~R.~Slatyer,
  [arXiv:1109.6322 [astro-ph.CO]].

\bibitem{Galli:2011rz}
  S.~Galli, F.~Iocco, G.~Bertone, A.~Melchiorri,
  Phys.\ Rev.\  {\bf D84 } (2011)  027302.
  
\bibitem{Jarosik:2010iu}
  N.~Jarosik, C.~L.~Bennett, J.~Dunkley, B.~Gold, M.~R.~Greason, M.~Halpern, R.~S.~Hill, G.~Hinshaw {\it et al.},
  Astrophys.\ J.\ Suppl.\  {\bf 192 } (2011)  14.
  
\bibitem{Trotta:2008qt}
  R.~Trotta,
  Contemp.\ Phys.\  {\bf 49} (2008) 71.

\bibitem{Trotta:2008bp}
  R.~Trotta, F.~Feroz, M.~P.~Hobson, L.~Roszkowski and R.~Ruiz de Austri,
  JHEP {\bf 0812} (2008) 024.
  
\bibitem{Roszkowski:2009ye}
  L.~Roszkowski, R.~Ruiz de Austri and R.~Trotta,
  Phys.\ Rev.\  D {\bf 82} (2010) 055003.
  
\bibitem{deAustri:2006pe}
  R.~R.~de Austri, R.~Trotta, L.~Roszkowski,
  JHEP {\bf 0605}, 002 (2006).
   
\bibitem{SuperBayeS}
  http://www.superbayes.org/
  
\bibitem{Feroz:2008xx}
  F.~Feroz, M.~P.~Hobson and M.~Bridges,
  arXiv:0809.3437 [astro-ph].

\bibitem{Arnowitt:2008bz}
  R.~L.~Arnowitt, B.~Dutta, A.~Gurrola, T.~Kamon, A.~Krislock and D.~Toback,
  Phys.\ Rev.\ Lett.\  {\bf 100} (2008) 231802.
  

\bibitem{Peebles:1968ja}
  P.~J.~E.~Peebles,
  Astrophys.\ J.\  {\bf 153} (1968) 1.

\bibitem{Zeldovich:1969en}
  Y.~B.~Zeldovich, V.~G.~Kurt and R.~A.~Sunyaev,
  J.\ Exp.\ Theor.\ Phys.\  {\bf 28} (1969) 146
  [Zh.\ Eksp.\ Teor.\ Fiz.\  {\bf 55} (1968) 278].


\bibitem{Brun:2011rg}
  P.~Brun,
  arXiv:1101.2745 [astro-ph.HE].

\bibitem{Consortium:2010bc}
  T.~C.~Consortium,
  arXiv:1008.3703 [astro-ph.IM].
  
\bibitem{Brun:2010ci}
  P.~Brun, E.~Moulin, J.~Diemand and J.~F.~Glicenstein,
  Phys.\ Rev.\  D {\bf 83} (2011) 015003.

\bibitem{Glicenstein}
  http://indico.in2p3.fr/contributionDisplay.py?sessionId=\-39\&contribId=88\&confId=1697


\bibitem{Bernloehr:2008xd}
  K.~Bernloehr,
  AIP Conf.\ Proc.\  {\bf 1085} (2009) 874.

\bibitem{Abdo:2010ex}
  A.~A.~Abdo, M.~Ackermann, M.~Ajello, W.~B.~Atwood, L.~Baldini, J.~Ballet, G.~Barbiellini, D.~Bastieri {\it et al.},
  Astrophys.\ J.\  {\bf 712 } (2010)  147-158.

\bibitem{Gelmini:2010zh}
  G.~Gelmini, P.~Gondolo,
  In *Bertone, G. (ed.): Particle dark matter* 121-141
  [arXiv:1009.3690 [astro-ph.CO]].

\end{thebibliography}
\end{document}